\documentclass[mathleft]{an}
\usepackage{graphicx}
\usepackage{amssymb}
\usepackage{amsmath}
\usepackage{latexsym}
\usepackage{times}
\usepackage{natbib}
\bibpunct{(}{)}{;}{a}{}{}

\begin{document}
\Pagespan{789}{}
\Yearpublication{2010}
\Yearsubmission{2010}
\Month{11}  
\Volume{999}  
\Issue{88} 
% \DOI{This.is/not.aDOI}% 

\title{Stellar Archaeology --- Exploring the Universe with Metal-Poor Stars}
\subtitle{\large Ludwig Biermann Award Lecture 2009}

\author{A. Frebel\thanks{Corresponding author:  \email{afrebel@cfa.harvard.edu}}}
\titlerunning{Exploring the Universe with metal-poor stars}
\authorrunning{Frebel}
\institute{
Harvard-Smithsonian Center for Astrophysics,
60 Garden St., MS-20,
Cambridge, MA 02138, USA}

\received{xx}
\accepted{xx}
\publonline{xx}


\abstract{The abundance patterns of the most metal-poor stars in the
Galactic halo and small dwarf galaxies provide us with a wealth of
information about the early Universe. In particular, these old
survivors allow us to study the nature of the first stars and
supernovae, the relevant nucleosynthesis processes responsible for the
formation and evolution of the elements, early star- and galaxy
formation processes, as well as the assembly process of the stellar
halo from dwarf galaxies a long time ago.  This review presents the
current state of the field of ``stellar archaeology'' -- the diverse
use of metal-poor stars to explore the high-redshift Universe and its
constituents.  In particular, the conditions for early star formation
are discussed, how these ultimately led to a chemical evolution, and
what the role of the most iron-poor stars is for learning about
Population\,III supernovae yields. Rapid neutron-capture signatures
found in metal-poor stars can be used to obtain stellar ages, but also
to constrain this complex nucleosynthesis process with observational
measurements.  Moreover, chemical abundances of extremely metal-poor
stars in different types of dwarf galaxies can be used to infer
details on the formation scenario of the halo. and the role of dwarf
galaxies as Galactic building blocks. I conclude with an outlook as to
where this field may be heading within the next decade. A table of
$\sim1000$ metal-poor stars and their abundances as collected from the
literature is provided in electronic format.}

\maketitle

%%%%%%%%%%%%%%%%%%%%%%%%%%%%%%%%%%%%%%%%%%%%%%
\section{Introduction}

As Carl Sagan once remarked, \textit{If you wish to make an apple pie
from scratch, you must first create the Universe}. An apple contains
at least 16 different
elements\footnote{http://www.food-allergens.de/symposium-vol1(3)/data/apple/apple-composition.htm},
and the human body is even more complex, having at least trace amounts
of nearly 30
elements\footnote{http://chemistry.about.com/cs/howthingswork/f/blbodyelements.htm},
all owing to a 14-billion year long manufacturing process called
cosmic chemical evolution.  Thus, the basis of chemically complex and
challenging undertakings such as cooking and baking, not to mention
the nature of life, will ultimately be gained through an understanding
of the formation of the elements that comprise organic material. It is
thus important to examine how the constituents of an apple, and by
extension the stuff of life and the visible Universe were created:
baryonic matter in the form of elements heavier than primordial
hydrogen and helium. In this review I will describe how the chemical
abundances observed in the most metal-poor stars are employed to
unravel a variety of details about the young Universe, such as early
star formation, nucleosynthesis in stars and supernovae (SNe), and the
formation process(es) of the Galactic halo. This concept is often
called ``stellar archaeology'' and is frequently used to address a
number of important, outstanding questions:

%-------------------------------
\begin{itemize}
%\vspace{-0.3cm}
\item What is the nature of Pop\,III stars? Are the yields of
  the first SNe different from today's? Can we find the
  signatures of theorized pair-instability SNe?
\item What drove early star formation?  How/where did the first
low-mass stars and the first galaxies form?
\item What are the main nucleosynthesis processes and sites that are
  responsible for forming the elements from the Big Bang until today?
\item How did chemical evolution proceed? How do stellar chemistry and
 halo kinematics correlate? How can we use abundances to learn about
the halo formation process?
\item Was the old halo built from accreted satellites? Can we identify
accreted dwarf galaxies in the halo? Did the first stars form in dwarf
galaxies?
\end{itemize}

Even though some of these question appear to refer to completely
unrelated topics, metal-poor stars do provide us with a powerful tool
to study a very broad range of astrophysical issues ranging from
nuclear astrophysics to early galaxy formation.  Metal-poor stars
represent the local \linebreak equivalent to the high-redshift Universe, and thus
provide a unique tool to address a wide range of near and far-field
cosmological topics. 

Each section of this review discusses one of those topics and can be
read independently, although it is advisable to also peruse
Section~2. It sets the overall stage by introducing the first stars,
early low-mass star formation and whether stellar archaeology is a
valid concept to study the high-redshift Universe.  Section~3 then discusses
the most iron-poor stars and their connection to early SNe, whereas in
Section~4, the observed signatures of neutron-capture nucleosynthesis
are presented. In particular, the r-process and nucleo-chronometry are
discussed. Section~5 deals with the metal-poor stars recently found in
small dwarf galaxies and what we can learn from their chemical
abundances about the early formation process of the Galactic halo.
Conclusions and an outlook at given in Section~6.

Since there exist a large range of metal-poor stars in terms of their
metallicities and chemical signatures, \citet{ARAA} suggested a
classification scheme. In this review I will make extensive use of
their term ``extremely metal-poor stars'', referring to stars with
$\mbox{[Fe/H]}<-3.0$.  ``Ultra metal-poor'' then refers to
$\mbox{[Fe/H]}<-4.0$, and ``hyper metal-poor'' to
$\mbox{[Fe/H]}<-5.0$.  This already shows that the main metallicity
indicator used to determine any stellar metallicity is the iron
abundance, [Fe/H], which is defined as \mbox{[A/B]}$ =
\log_{10}(N_{\rm A}/N_{\rm B})_\star - \log_{10}(N_{\rm A}/N_{\rm
B})_\odot$ for the number N of atoms of elements A and B, and $\odot$
refers to the Sun. With few exceptions, [Fe/H] traces the overall
metallicity of the objects fairly well.

\section{The Early Universe}

\subsection{The First Stars}

According to cosmological simulations that are based on the $\Lambda$
cold dark matter model of hierarchical structure growth in the
Universe, the first stars formed in small minihalos some few hundred
million years after the Big Bang. Due to the lack of cooling agents in
the primordial gas, significant fragmentation was largely suppressed
so that these first objects were very massive (of the order to
$\sim100 $\,M$_{\odot}$; e.g, \citealt{brommARAA} and references
therein). This is in contrast to low-mass stars dominating today's mass
function. These objects are referred to as Population\,III (Pop\,III)
as they formed from metal-free gas.

These objects soon exploded as SNe to either collapse into
black holes (progenitor masses of $25<M_{\odot}<140$ and
$M_{\odot}>260$) or to die as energetic pair-instability SNe 
(PISN; $140<M_{\odot}<260$; \citealt{heger2002}). During their deaths, these
objects provided vast amounts of ionizing radiation (and some of the
first metals in the case of the PISNe) that changed the conditions of
the surrounding material for subsequent star formation even in
neighboring minihalos.  Hence the second generation of stars might
have been less massive (M$_{\star}\sim10\,$M$_{\odot}$). Partially
ionized gas supports the formation of the H$_{2}$, and then the HD
molecule which in turn facilitates more effective cooling than what is
possible in neutral gas. Also, any metals or dust grains left behind
from PISNe would have similar cooling effects. This may then have led
to the first more regular metal-producing SNe, although not all
higher mass SNe must necessarily end in black hole
formation. \citet{UmedaNomotoNature} suggested that some
25\,M$_{\odot}$ stars undergo only a partial fallback, so that some of the
newly created metals get ejected into the surrounding gas. By that
time, most likely enough metals were present to ensure sufficient gas
fragmentation to allow for low-mass ($<$1\,M$_{\odot}$) star
formation.  Those stars that formed from any metal-enriched material
are referred to as Population\,II (Pop\,II) stars. More metal-rich
stars like the Sun that formed in a much more metal-rich Universe
are called Population\,I.

The concept of stellar archaeology entails that the most extreme
versions of Pop\,II stars preserve in their surface composition the
individual SN yields of previous Pop\,III stars. Studying those
``chemical fingerprints'' in the oldest, most metal-poor stars can
thus reveal a great deal about the first nucleosynthesis events in the
Universe. Indeed, several metal-poor star abundance patterns have been
fitted with calculated Pop\,III SN yields.  Moreover, one may also
seek to find a PISN signature (a pronounced odd-even abundance
signature) in metal-poor stars. This has, however, not yet occurred.

\subsection{Early low-mass star formation and the connection to carbon-enhanced metal-poor stars}\label{sec_arch}

Early Pop\,II stars began to form from the enriched material
left behind by the first stars. The actual formation process of these
initial low-mass (M $\le$ 0.8 M$_{\odot}$) Pop\,II stars
(i.e. the most metal-poor stars) that live longer than a Hubble time,
is not well-understood so far. Ideas for the required cooling
processes necessary to induce sufficient fragmentation of the
near-primordial gas include cooling through metal enrichment
(``critical metallicity'') or dust, cooling based on enhanced molecule
formation due to ionization of the gas, as well as more complex
effects such as turbulence and magnetic fields \citep{bromm09}.

Fine-structure line cooling through neutral carbon and singly-ionized
oxygen has been suggested as a main cooling agent facilitating
low-mass star formation \citep{brommnature}.  These elements were
likely produced in vast quantities in Pop\,III objects
(e.g. \citealt{meynet06}). Gas fragmentation is then induced once a
critical metallicity of the interstellar medium (ISM) is achieved.
The existence of such a critical metallicity can be probed with large
numbers of \textit{carbon and oxygen-poor} metal-poor stars.  Frebel
et al. (2007b) developed an ``observer-friendly'' description of the
critical metallicity that incorporates the observed C and/or O stellar
abundances; $D_{\rm trans}={\rm log} (10^{{\rm [C/H]}} + 0.3 \times
10^{{\rm [O/H]}})\ge-3.5$.  Any low-mass stars still observable today
then has to have C and/or O abundances above the threshold of $D_{\rm
trans}=-3.5$ (see Figure~1 in \citealt{dtrans}). At metallicities of
$\mbox{[Fe/H]}\gtrsim-3.5$, most stars have C and/or O abundances that
are above the threshold since they follow the solar C and O abundances
simply scaled down to their respective Fe values. Naturally, this
metallicity range is not suitable for directly probing the first
low-mass stars. Below $\mbox{[Fe/H]}\sim-3.5$, however, the observed C
and/or O levels must be higher than the Fe-scaled solar abundances to
be above the critical metallicity. Indeed, none of the known
lowest-metallicity stars has a $D_{trans}$ below the critical \linebreak  value,
consistent with this cooling theory. Some stars, however, have values
very close to $D_{\rm trans}=-3.5$. \linebreak  HE~0557$-$4840, at
$\mbox{[Fe/H]}=-4.75$ \citep{he0557}, falls just onto the critical
limit (M. Bessell 2009, priv. \linebreak comm.). A star in the ultra-faint dwarf
galaxy Bo\"otes\,I has $D_{\rm trans}=-3.2$ (at $\mbox{[Fe/H]}=-3.7$;
and assuming that its oxygen abundance is twice that of carbon). An
interesting case is also the most metal-poor star in the classical
dwarf galaxy Sculptor, which has an upper limit of carbon of
$\mbox{[C/H]}<-3.6$ at $\mbox{[Fe/H]}=-3.8$ \citep{scl}. Despite some
still required up-correction of carbon to account for atmospheric
carbon-depletion of this cool giant, the star could potentially posses
a sub-critical $D_{\rm trans}$ value.

Overall, more such ``borderline'' examples are crucial to test for the
existence of a critical metallicity. If fine-structure line cooling
were the dominant process for low-mass star formation, two important
consequences would follow: 1) Future stars to be found with
$\mbox{[Fe/H]}\lesssim-4.0$ are predicted to have these significant C
and/or O overabundances with respect to Fe.  2) The so-far unexplained
large fraction of metal-poor objects that have large overabundances of
carbon with respect to iron ($\mbox{[C/Fe]}>1.0$) may reflect an
important physical cause.  About 20\% of metal-poor stars with
$\mbox{Fe/H}\lesssim-2.5$ exhibit this behavior
(e.g. \citealt{ARAA}). Moreover, at the lowest metallicities, this
fraction is even higher. All three stars with $\mbox{[Fe/H]}<-4.0$ are
extremely C-rich, well in line with the prediction of the line cooling
theory.

\begin{figure*}  [!]
\begin{center}
\includegraphics[width=16.5cm,clip=true,bbllx=65,bblly=423,bburx=528,bbury=655]{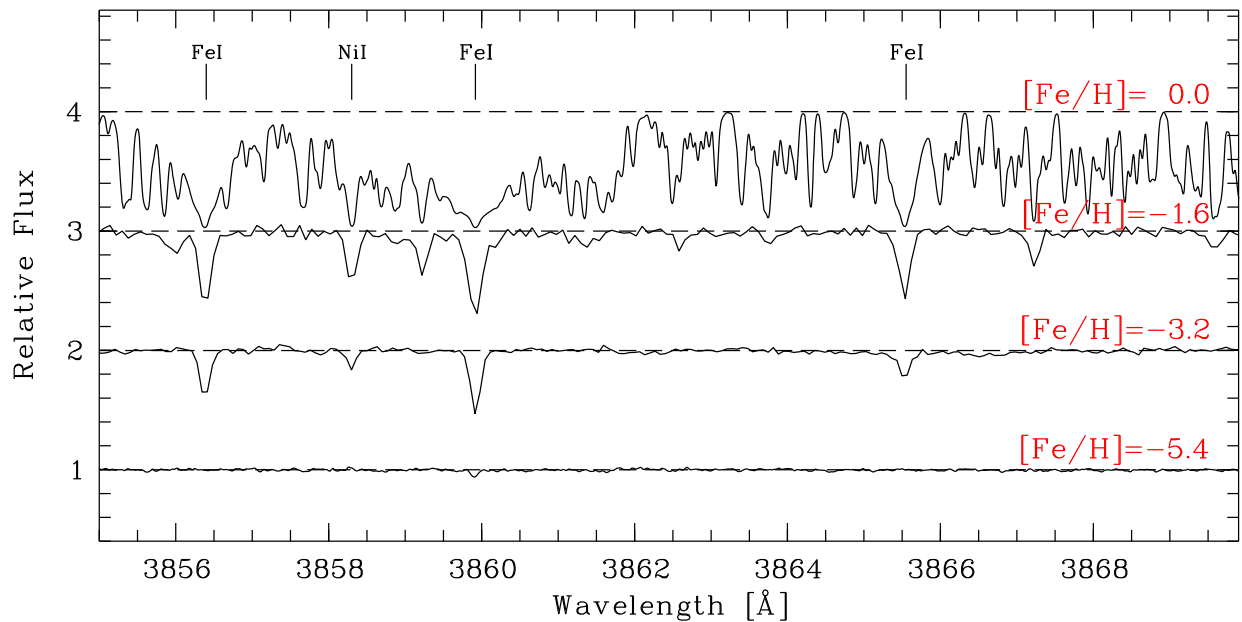}
  \caption{\label{spec_comp} Spectral comparison of stars in the
  main-sequence turn-off region with different metallicities. Several
  absorption lines are marked. The variations in line strength reflect
  the different metallicities. From top to bottom: Sun with
  $\mbox{[Fe/H]}=0.0$, G66-30 $\mbox{[Fe/H]}=-1.6$ \citep{norris_emp3}, G64-12
  $\mbox{[Fe/H]}=-3.2$ \citep{HE1327_Nature}, and HE1327-2326 $\mbox{[Fe/H]}=-5.4$ \citep{HE1327_Nature}.  }
\end{center}
  \end{figure*}

It should also be mentioned that cooling through dust grains might
have been responsible for the transition from Pop\,III to Pop\,II star
formation. Dust created during the first SNe explosions or mass loss
during the evolution of Pop\,III stars may induce fragmentation
processes (e.g., \citealt{schneider06}) that lead to the formation of
subsolar-mass stars. The critical metallicity in this scenario is a
few orders of magnitude below that of C and O line cooling.  If some
\linebreak metal-poor stars are found to be below $D_{\rm trans}=-3.5$, they may
still be consistent with the critical value set by dust cooling.

\subsection{Validating Stellar Archaeology}

Stellar archaeology is based on long-lived low-mass metal-poor
main-sequence and giant stars whose chemical abundances are thought to
reflect the composition of the ISM during their formation. A vital
assumption is that the stellar surface compositions have not been
significantly altered by any internal mixing processes given that
these stars are fairly unevolved despite their old age. But are there
other means by which the surface composition could be modified?
Accretion of interstellar matter while a star orbits in the Galaxy for
$\sim10$\,Gyr has long been suggested as a mechanism to affect the
observed abundance patterns. \citet{iben1983} calculated a basic
"pollution limit" of $\mbox{[Fe/H]}=-5.7$ based on Bondi-Hoyle
accretion. He predicted that no stars with Fe abundances below this
value could be found since they would have accreted too much enriched
material.

Assuming that stars with such low-metallicities exist \linebreak (for example
low-mass Pop\,III stars if the IMF was Salpeter-like, and not
top-heavy),  significant amounts of interstellar accretion could
masquerade the primordial abundances of those putative low-mass
Pop\,III stars. Analogously, stars with very low abundances, say
 $\mbox{[Fe/H]}<-5.0$, could principally be affected also.
\citet{poll} carried out a kinematic analysis of a sample of
metal-poor stars to assess their potential accretion histories over
the past 10\,Gyr in a Milky Way-like potential. The amount of accreted
Fe was calculated based on the total accreted material over
10\,Gyr. The overall chemical evolution with time was taken into
account assuming the ISM to have scaled solar
abundances.

The stellar abundances were found to be little affected by
accretion. The calculated, ``accreted abundances'' were often lower than
the observed measurements by several orders of magnitude.  Generally,
this confirms that accretion does not significantly alter the observed
abundance patterns, even in an extreme case in which a star moves once
through a very large, dense cloud.  The concept of stellar archaeology
is thus viable. Nevertheless, since there is a large accretion
dependency on the space velocity it becomes obvious that kinematic
information is vital for the identification of the lowest-metallicity
stars in the Milky Way and the interpretation of their abundances.

\subsection{The metallicity distribution function}

Large numbers of Galactic metal-poor stars found in \linebreak
objective-prism surveys in both hemispheres have provided great
insight into the history and evolution of our Galaxy (e.g.,
\citealt{ARAA}). However, there are still only very few stars known
($\lesssim20$) with metallicities below \linebreak $\mbox{[Fe/H]}<-3.5$.
Recently, \citet{schoerck} presented a new metallicity distribution
function (MDF) for halo stars that is corrected for various selection
effects and other biases. The number of known metal-poor stars
declines significantly with decreasing metallicity (below
$\mbox{[Fe/H]}<-2.0$) as illustrated in their Figures~10 and 18. The
new \linebreak bias-corrected MDF also shows how rare metal-poor stars really
are, but also that past targeted (``biased'') searches for metal-poor
stars have been extremely successful at identifying these objects.
The biggest achievements in terms of the most iron-deficient stars was
the push to a significantly lower stellar metallicity $\mbox{[Fe/H]}$
almost a decade ago: From a longstanding $\mbox{[Fe/H]}=-4.0$ (CD
$-38^{\circ}$ 245; \citealt{cd38}) to $\mbox{[Fe/H]}=-5.2$
(HE~0107$-$5240; \citealt{HE0107_Nature}\footnote{Applying the same
non-LTE correction to the Fe\,I abundances of HE~0107$-$5240 and
HE~1327$-$2326 leads to a final abundance of $\mbox{[Fe/H]}=-5.2$ for
HE~0107$-$5240.}), and more recently, down to $\mbox{[Fe/H]}=-5.4$
(HE~1327$-$2326; \citealt{HE1327_Nature}). Overall, only three stars
are known with iron abundances of $\mbox{[Fe/H]}<-4.0$. The recently
discovered star HE~0557$-$4840 \citep{he0557} with
$\mbox{[Fe/H]}<-4.8$ bridges the gap between \linebreak $\mbox{[Fe/H]}=-4.0$ and
the two hyper Fe-poor objects. Objects in the very tail of
the MDF provide a unique observational window onto the time very
shortly after the Big Bang. They provide key insights into the very
beginning of Galactic chemical evolution.

To illustrate the progression of metallicity from metal-rich to the
most metal-poor stars, Figure~\ref{spec_comp} shows spectra around the
strongest optical Fe line at 3860\,{\AA} of the Sun and four other
metal-poor main-sequence stars. The number of atomic absorption lines
detectable in the spectra decreases with increasing
metal-deficiency. In HE~1327$-$2326, only the intrinsically strongest
metal lines remain observable, \linebreak and these are extremely weak. If a
main-sequence star with even lower Fe value was discovered, no Fe
lines would be measurable anymore. In the case of a giant, the lines
would be somewhat stronger due to its cooler temperature and thus
allow for the discovery of a $\mbox{[Fe/H]}\lesssim-6$ object.

%%%%%%%%%%%%%%%%%%%%%%%%%%%%%%%%%%%%%%%%%%%%%%
\section{Studying the Early Universe with Metal-Poor Stars}

\subsection{Searching for the Most Metal-Poor Stars}

Over the past two decades, the quest to find the most metal-poor stars
to study the chemical evolution of the Galaxy led to a significant
number of stars with metallicities down to $\mbox{[Fe/H]}\sim-4.0$
(see \citealt{ARAA} for a more detailed review). Those stars were
initially selected as candidates from a large survey, such as the HK
survey \citep{BPSII} and the Hamburg/ESO survey \citep{hespaperI}.  A
large survey is required to provide numerous low-resolution spectra to
search for weak-lined stellar candidates. Those spectra have to cover
the strong Ca\,II\,K line at 3933\,{\AA} because the strength of this
line indicates the metallicity of the star, and can be measured in
low-quality spectra. If this line is sufficiently weak as a function
of the star's estimated effective temperature, an object is selected
as a candidate metal-poor star. For all candidates, medium-resolution
spectra ($R\sim2000$) are required to more accurately determine the
\linebreak Ca\,II\,K line strength for a more robust estimate for the
Fe abundance. This line is still the best indicator for the overall
metallicity [Fe/H] of a metal-poor star in such spectra. In the Sloan
Digital Sky Survey and LAMOST survey, the survey spectra themselves
are already of medium-resolution allowing for a quicker and more
direct search for metal-poor stars. To confirm the metallicity, and to
measure elemental abundances from their respective absorption lines
besides that of iron, high-resolution optical spectroscopy is
required. Only then the various elements become accessible for studying
the chemical evolution of the Galaxy. Those elements include carbon,
magnesium, calcium, titanium, nickel, \linebreak strontium, and
barium, and trace different enrichment mechanisms, events and
timescales. Abundance ratios [X/Fe] as a function of [Fe/H] can then
be derived for the lighter elements ($Z<30$) and neutron-capture
elements ($Z>38$). The final number of elements hereby depends on the
type of metal-poor star, the wavelength coverage of the data, and the
data quality itself.

\subsection{Chemical Evolution of the Galaxy}

Generally, there are several main groups of elements observed in
metal-poor stars, with each group having a common, main production
mechansism; 1) $\alpha$-elements (e.g. Mg, Ca, Ti) are produced
through $\alpha$-capture during various burning stages of late stellar
evolution, before and during SN explosions. These yields appear very
robust with respect to parameters such as mass and explosion energy 2)
Fe-peak elements (23$<Z<30$) are synthesized in a host of different
nucleosynthesis processes before and during SN explosions such as
radioactive decay of heavier nuclei or direct synthesis in explosive
burning stages, neutron-capture onto lower-mass Fe-peak elements
during helium and later burning stages and alpha-rich freeze-out
processes. Their yields also depend on the explosion energy; 3) Light
and heavy neutron-capture elements ($Z>38$) are either produced in the
slow (s-) process occurring in thermally pulsing AGB stars (and then
transferred to binary companions or deposited into the ISM through
stellar winds) or in the rapid (r-) process most likely occurring in
core-collapse SN explosions. For more details on SN nucleosynthesis
see e.g., \citet{woosley_weaver_1995}.

\begin{figure}  [!htb]
\includegraphics[height=11.7cm,clip=true,bbllx=43,bblly=40,bburx=550,bbury=752]{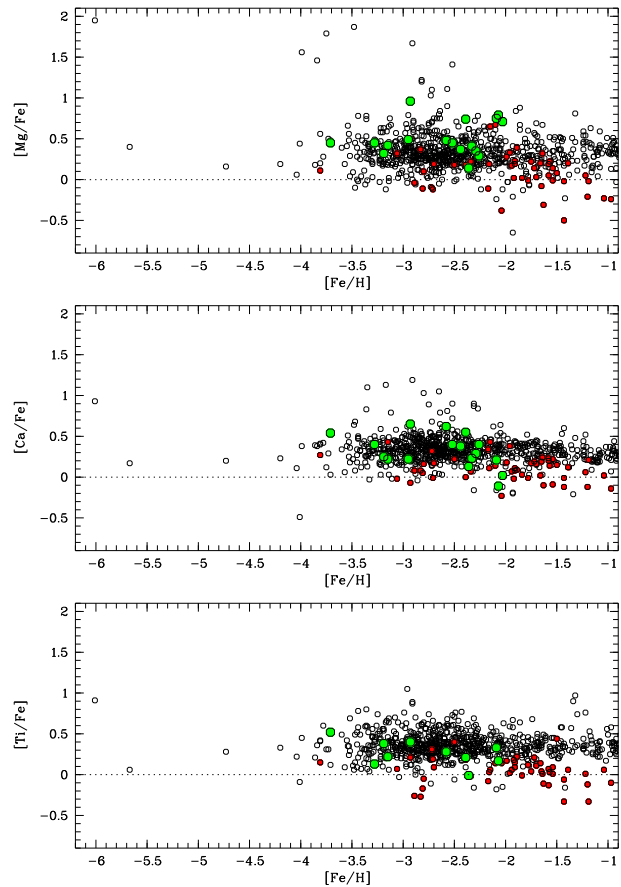}
  \caption{ \label{alphas} Light element abundance trends of Mg, Ca,
  and Ti. Black open circles represent halo stars, red fille circles
  are stars in the classical dwarf galaxies, and green filled circles
  show stars in the ultra-faint dwarf galaxies. Both the x- and y-axis
  have the same scale for easy comparisons. Only stars wtih
  high-resolution abundance analyses are shown. The abundance data can
  be found in an electronic format. The scatter in the data likely
  reflects systematic differences between literature studies. Assuming
  this, systematic uncertainties in abundance analyses may be around
  $\sim0.3$\,dex.}
  \end{figure}

The $\alpha$-element abundances in metal-poor halo stars with
$\mbox{[Fe/H]}\sim-1.5$ are enhanced by $\sim0.4$\,dex with respect to
Fe (see Figure~\ref{alphas}). This reflects a typical core-collapse SN
signature because at later times (in chemical space at about
$\mbox{[Fe/H]}\sim-1.5$) the onset of SN Ia provides a significant
contribution to the overall Galactic Fe inventory. As a consequence,
the \mbox{[$\alpha$/Fe]} ratio decreases down to the solar value at
$\mbox{[Fe/H]}\sim0.0$.  The general uniformity of light element
abundance trends down to $\mbox{[Fe/H]}\sim-4.0$ led to the conclusion
that the ISM must have been already well-mixed at very early times
\citep{cayrel2004}. Otherwise it would be hard to understand why so
many of the most metal-poor stars have almost identical abundance
patterns. However, despite the well-defined abundance trends, some
stars, particularly those in the lowest metallicity regime show
significant deviations. Some stars have been found to be very
$\alpha$-poor (e.g. \citealt{ivans_alphapoor}) and others are strongly
overabundant in Mg and Si (e.g., \citealt{aoki_mg, HE1327_Nature}).

\begin{figure}  [!htb]
\includegraphics[height=11.7cm,clip=true,bbllx=43,bblly=40,bburx=550,bbury=752]{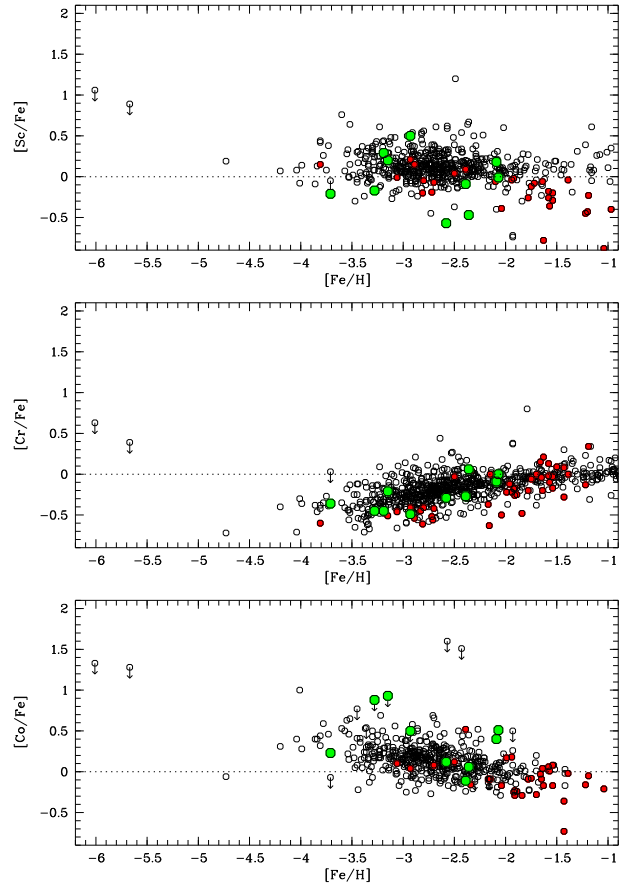}
  \caption{ \label{fepeak} Same as Figure~\ref{alphas}, but for the
  Fe-peak elements Sc, Cr, and Co.}
  \end{figure}

Among the Fe-peak elements, many have subsolar abundance trends at low
metallicity (e.g. [Cr,Mn/Fe]) which become solar-like as the
metallicity increases (see Figure~\ref{fepeak}). It is not clear
whether these large underabundances are of cosmic origin or have to be
attributed to modelling effects such as that of non-LTE
(\citealt{sobeck07,bergemann_mn}). Trends of other elements are somewhat
overabundant at low metallicity (Co) or relatively unchanged
throughout (Sc, Ni).  All elements with $Z<30$ hereby have relatively tight
abundance trends.

How can all those signatures be understood? The observed abundances of
the most metal-poor stars with ``classical'' halo signatures have
successfully been reproduced with Pop\,III SN yields.
\citet{tominaga07_b} model the averaged abundance pattern of four
non-carbon-enriched stars with $-4.2<\mbox{[Fe/H]}<-3.5$ with the
elemental yields of a massive, energetic ($\sim30-50$\,M$_{\odot}$)
Pop\,III hypernova. The abundances can also be fitted with
integrated yields of Pop\,III SNe \citet{heger_woosley08}.
Special types of SNe or unusual nucleosynthesis yields can then be
considered for stars with chemically peculiar abundance patterns. It
is, however, often difficult to explain the entire abundance
pattern. Additional metal-poor stars as well as a better understanding
of the explosion mechansism and the impact of the initial conditions on
SNe yields are required to arrive at a more solid picture of exactly
how metal-poor stars reflect early SNe yields.

\begin{figure*}  [!htb]
\includegraphics[width=17.1cm,clip=true,bbllx=32,bblly=298,bburx=545,bbury=499]{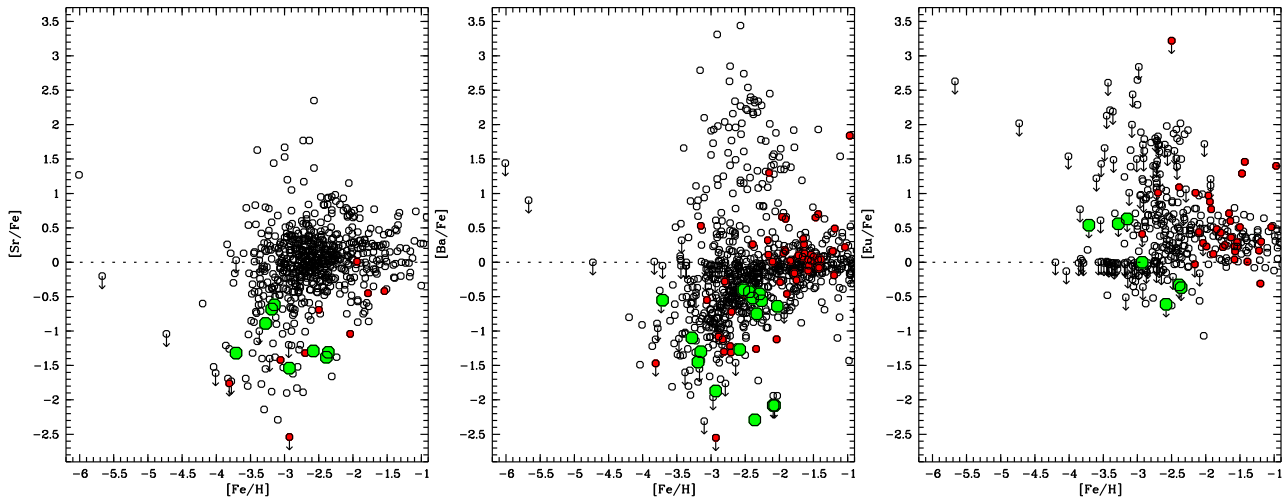}
  \caption{ \label{ncap} Same as Figure~\ref{alphas}, but for the
  neutron-capture elements Sr, Ba, and Eu. For these elements, the
  scatter is much beyond systemtic differences between individual
  studies and thus indicates a cosmic origin.}
  \end{figure*}

On the contrary, the abundances of the neutron-capture elements in
metal-poor stars are ``all over the place''.  Sr has an extremely
large scatter ($\sim 3$\,dex). This indicates that different
nucleosynthetic processes must have contributed to its Galactic
inventory, or that neutron-capture yields are very
environmentally-sensitive. Ba has even more scatter at
$\mbox{[Fe/H]}\sim-3.0$ (see Figure~\ref{ncap}). Other heavier
neutron-capture elements, such as Eu, have somewhat less scatter but this may
be due to their generally weak and usuallly more difficult to detect
absorption lines. What is apparent, though, is that at the lowest
metallicities, core-collapse SNe must have dominated the chemical evolution
(below $\mbox{[Fe/H]}=-3.0$). \linebreak Hence, the r-process is
responsible for the neutron-capture elements at this early time. The
s-process contribution occurred at somewhat later times, driven by the
evolutionary timescales of stars with $\sim2$-8\,M$_{\odot}$ to become
AGB stars.

To illustrate the extent of the chemical evolution of the Galaxy we
have collected abundance data of metal-poor \linebreak stars from the
literature (Figures~\ref{alphas}, \ref{fepeak} and \ref{ncap}). All
abundances [X/Fe] have been recalculated with the latest solar abundances of
\citet{asplund09}. In the many cases where a star has been studied
more than once, the study with the most elements (i.e. likely with the
highest quality data) was usually picked. No other processing of the
literature data has been done. Hence, systematic differences between
different studies remain. The tables containing all the abundances are
easily accessible from the website of the Astronomische Nachrichten
(www.aip.de/AN/).  Abundance plots for all elements, the corresponding
tables and more explanations are  available at \\
www.cfa.harvard.edu/$\sim$afrebel/abundances/abund.html References are assigned
to each set of abundances in the table. They are also alphabetically
listed in the Appendix. For many stars, a key tag has been assigned to
each star corresponding to its chemical properties. The keys are
listed on the website as well. A similar collection of data can be
found in the SAGA database \citep{saga} which is independent of the
current collection.

\subsection{The most iron-poor stars}

The first star with a record-low iron abundance was found in 2001. The
faint ($V=15.2$) red giant HE~0107$-$5240 has $\mbox{[Fe/H]}=-5.2$
\citep{HE0107_Nature}. In 2004, the bright ($V=13.5$) subgiant
HE~1327$-$2326 was discovered \citep{HE1327_Nature,
Aokihe1327}. HE~1327$-$2327 has an even lower iron abundance of
$\mbox{[Fe/H]}=-5.4$. This value corresponds to $\sim1/250,000$ of the
solar iron abundance. Interestingly, the entire mass of iron in
HE~1327$-$2326 is actually 100 times less than that in the Earth's
core. At the same time the star is of course of the order of a million
times more massive than the Earth.  A third star with
$\mbox{[Fe/H]}=-4.75$ \citep{he0557} was found in 2006. The
metallicity of the giant HE~0557$-$4840 is in between the two $<-5.0$
stars and the next most metal-poor stars are
$\mbox{[Fe/H]}\sim-4.0$. Hence it sits right in the previously claimed
``metallicity gap'' (between $\mbox{[Fe/H]}\sim-4.0$ and
$\mbox{[Fe/H]}\sim-5.0$; e.g. \citealt{shigeyama}) showing that the
scarcity of stars below $\mbox{[Fe/H]}=-4.0$ is not a physical cause
but a merely an observational incompleteness. All three objects were
found in the Hamburg/ESO survey making it the so far most successful
database for metal-poor stars.

The most striking features in both $<-5.0$ stars are the extremely
large overabundances of CNO elements. \linebreak HE~0557$-$4840 partly
shares this signature by also having a fairly large [C/Fe] ratio.
Other elemental ratios [X/Fe] are somewhat enhanced in HE~1327$-$2327
with respect to the stars with $-4.0<\mbox{[Fe/H]}<-2.5$, but less so
for the two giants. No neutron-capture element is detected in
HE~0107$-$5240 or HE~0557$-$4840, whereas, unexpectedly, Sr is
observed in HE~1327$-$2326. The Sr may originate from the
neutrino-induced $\nu p$-process operating in SN explosions
\citep{froehlich}.  Despite expectations, Li \linebreak could not be
detected in the relatively unevolved subgiant HE~1327$-$2326. The
upper limit is $\log\epsilon ({\rm Li})<1.6$, where $\log\epsilon
({\rm A})$ = $\log_{10}(N_{\rm A}/N_{\rm H})$ + 12. This is
surprising, given that the primordial Li abundance is often inferred
from similarly unevolved metal-poor stars
\citep{ryan_postprim}. Furthermore, the upper limit found from
HE~1327$-$2326, however, strongly contradicts the WMAP value
($\log\epsilon ({\rm Li})=2.6$) from the baryon-to-photon ratio
\citep{WMAP}. This may indicates that the star formed from extremely
Li-poor material.

HE~0107$-$5240 and HE~1327$-$2326 immediately became benchmark objects
to constrain various theoretical \linebreak studies of the early
Universe, such as the formation of the first stars (e.g.,
\citealt{yoshida06}), the chemical evolution of the early ISM (e.g.,
\citealt{karlsson2005}) or calculations of Pop\,III SN yields. Their
highly individual abundance patterns have been successfully reproduced
by several different SNe scenarios. This makes HE~0107$-$5240 and
HE~1327$-$2326 early, extreme Pop\,II stars that possibly
display the ``fingerprint'' of just one Pop\,III
SN. \citet{UmedaNomotoNature} first matched the yields of a faint
25\,M$_{\odot}$ SN that underwent a mixing and fallback process to the
observed abundances of HE~0107$-$5240.  To achieve a simultaneous
enrichment of a lot of C and only little Fe, large parts of the
Fe-rich SN ejecta have to fall back onto the newly created black
hole. Using yields from a SN with similar explosion energy and mass
cut, \citet{iwamoto_science} then reproduced the abundance pattern of
HE~1327$-$2326 also. Trying to fit the observed stellar abundances,
\citet{heger_woosley08} are employing an entire grid of Pop\,III SN
yields to search for the best match to the data. A similar progenitor
mass range as the \citet{UmedaNomotoNature} 25\,M$_{\odot}$ was found
to be the best match to have provided the elemental abundances to the
ISM from which these Pop\,II stars formed.  \citet{meynet2005}
explored the influence of stellar rotation on elemental yields of
60\,M$_{\odot}$ near-zero-metallicity SNe.  The stellar mass loss rate
of rotating massive Pop\,III stars qualitatively reproduces the CNO
abundances observed in HE~1327$-$2326 and other carbon-rich metal-poor stars.

\citet{limongi_he0107} were able to reproduce the
abundances of HE~0107$-$5240 through pollution of the birth cloud by
at least two SNe. \citet{suda} proposed that the abundances of
HE~0107$-$5240 would originate from a mass transfer of CNO elements
from a postulated companion, and from accretion of heavy elements from
the ISM. However, neither  HE~0107$-$5240 nor
HE~1327$-$2326 show radial velocity variations that would indicate
binarity (R. Lunnan in preparation).

%%%%%%%%%%%%%%%%%%%%%%%%%%%%%%%%%%%%%%%%%%%%%%
\section{Neutron-Capture Nucleosynthesis Observed in Metal-Poor Stars}

All elements heavier than the Fe-peak are created through
neutron-capture processes in stars during stellar evolution (via AGB
nucleosynthesis; slow $s$-process) and SN explosions (via the
rapid $r$-process). We here summarize the most important constraints
on the various neutron-capture processes provided by different types of
metal-poor stars. A short review of r-process nucleosynthesis as observed
in metal-poor stars can also be found in \citet{frebel_nic10}. For a
detailed review on how neutron-capture processes drove the
chemical evolution of the Galaxy, we refer the interested reader to
\citet{sneden_araa}.

\subsection{The r-Process Signature}
The r-process occurs when a seed nucleus (e.g., a Fe or C nucleus) is
under intensive neutron bombardment. In particular, when the
neutron-capture rate exceeds that of the $\beta$-decay,
large numbers of neutrons can be captured quickly to build up even the
heaviest nuclei in the periodic table ($38<Z\,<\,92$). The
astrophysical site(s) that can accommodate the required extreme
environment for this process have not yet been clearly
identified. Neutron-driven winds emerging from a proto-neutron-star
which formed after a SN explosion (perhaps with a progenitor of
$8-10$\,M$_{\odot}$) seem to be promising locations
\citep{qian_wasserburg03}. Neutron-star mergers have also been
considered, but their long evolutionary timescales make them
unsuitable as the primary r-process site in the early Galaxy
\citep{argast}. Not knowing the site, and hence, the initial
conditions, makes it difficult to study the r-process
theoretically. Furthermore, nuclear physics experiments on heavy,
exotic r-process nuclei are technically out of reach. This leaves few
experimental constraints on the production of the heaviest elements in
the Universe.

Fortunately, about $5\%$ of metal-poor stars with $\mbox{[Fe/H]}<-2.5$
contain a strong enhancement of neutron-capture element\footnote{Stars
with [r/Fe] $>1.0$; r represents the average abundance of elements
from the r-process.}  associated with the r-process \citep{ARAA}. In
those stars, we can observe the majority of elements in the periodic
table: the usual light, $\alpha$- and iron-peak, \textit{and} the full
set of optically available light ($38<Z<56$) and heavy ($56<Z<92$)
neutron-capture elements. These neutron-capture elements were not
produced in the observed metal-poor stars themselves, but in a
previous-generation SN explosion. The so-called r-process metal-poor
stars \linebreak formed from material that was chemically pre-enriched by
this SN.  We are thus able to study the ``chemical fingerprint'' of
individual r-process events that likely occurred shortly before the
formation of the observed star. Hence, the \linebreak r-process stars
fortuitously bring together astrophysics and nuclear physics by acting
as a ``cosmic laboratory'' for both fields of study, and provide crucial
experimental data on heavy-element production.

The picture that has been emerging from these observation is simple
and yet astounding.  The abundance patterns of the neutron-capture
elements particularly the heaviest ones with $Z>56$, agree extremely
well with each other in all r-process stars \citep{Snedenetal:1996,
johnson_bolte, Hilletal:2002,heresII, ivans06, he1523} and with the
scaled solar r-process pattern (e.g. \citealt{2000burris}). Overall,
this behavior suggests that the r-process is universal and leads to
the same elemental ratios in the early Universe as at much later
times, when the Sun was born\footnote{The Sun's r-process pattern can
be obtained by subtracting the calculated s-process component from the
measured total abundance pattern.}.  The universality offers a unique
constraint on any theoretical modeling of the r-process as the
end-product is ``known'' through the r-process stars.  It also enables
calculating elemental ratios involving radioactive elements such as
thorium and uranium that can be compared with observed measurements of
the left-over radioactive stellar material.  This, in turn, makes
possible nucleochronometric age dating of the oldest stars.

Among the lighter neutron-capture elements there are abundance
deviations between the scaled solar r-process \linebreak pattern and to some
extent among the individual stars. Since it is not clear if these
lighter elements are produced by a single r-process, additional new
processes were invoked in order to explain the entire spread
of observed neutron-capture abundances in this mass range (e.g.,
\citealt{travaglio, aoki05, wanajo05}). Suggestions include a ``weak''
r-process acting as a ``failed'' r-process only producing
elements with $Z<56$. An example of this process may the r-poor star
HD122563 \citep{honda07, honda06} showing a comparable, exclusive
light neutron-capture element enhancement. Unraveling the exact details
about the different r-processes should shed new light on the
astrophysical site(s) of the r-process.

A ``clean'' r-process signature can only be observed in those stars
that formed before the onset of the s-process
($\mbox{[Fe/H]}\sim-2.6$; \citealt{simmerer2004}) and at times when
the galactic inventory of lighter elements was still low so that
spectral contamination with metal lines is minimal. \linebreak Only the most
metal-poor stars can thus be examined for the r-process signature.
The first strongly enhanced r-process star was discovered more than a
decade ago in the HK survey, CS~22892-052 \citep{Snedenetal:1996}. A
second star was  found a few years thereafter, CS~31082-001
\citep{Cayreletal:2001}. Recognizing the need to find more such stars
to enable detailed studies of the r-process, \citet{heresI} initiated
a large campaign to observe metal-poor candidate stars from the main
Hamburg/ESO survey. It led to the discovery of several strongly
r-process-enhanced stars \linebreak \citep{heresII, hayek09} and dozens of mildly
enriched ones. Two more such stars were found more recently elsewhere
\citep{he1523,lai2008}.

The metallicity of all the strongly r-enriched stars is coincidentally
$\mbox{[Fe/H]}\sim-3.0$ or just above it. Regarding their
light-elements, no unusual abundances are found, and their
halo-typical enhancements in $\alpha$-elements suggest a core-collapse
SN the source responsible  for their overall elemental signature,
including the neutron-capture elements. This is in line with SNe being
the early contributors to galactic chemical evolution whereas AGB
enrichment  sets in only at somewhat later times due to their
evolutionary timescales and the presence of seed nuclei (i.e., from
Fe-peak elements created in previous SNe generations).

\subsection{Nucleo-Chronometry of the Oldest Stars}

The long-lived radioactive isotopes $^{232}$Th (half-life of
\linebreak $14$\,Gyr) and $^{238}$U ($4.5$\,Gyr) are suitable for measurements of
cosmic timescales. They also have transitions in the optical range so
that, in principle, Th and U abundances can be measured in stellar
spectra of r-enriched stars. However, suitable stars for these
challenging measurements are difficult to find: Cool metal-poor red
giants that exhibit very strong overabundances of r-process
elements. Their carbon abundances should be low (i.e. subsolar) since
CH features blend with many important neutron-capture lines (e.g. U,
Pb), rendering them unmeasurable in carbon-rich r-process stars such
as CS~22892-052. Moreover, the stars need to be bright (preferably
$V<13$) so that high-resolution data with very high $S/N$ can be
collected in reasonable observing times. Many neutron-capture features
are very weak and often partially blended and thus require
exceptionally high data quality. The two most important examples are
the extremely weak U line at 3859\,{\AA} and the even weaker Pb line
at 4057\,{\AA}. These two lines are the strongest optical transitions
of these element. Subject to the availability of a very suitable star, a
successful U measurement requires a $R>60,000$ spectrum with a $S/N$
of at least 350 per pixel at 3900\,{\AA}. A Pb measurement may be
attempted in a spectrum with $S/N\sim500$ at 4000\,{\AA}.

Fortunately, Th and several other stable r-process elements (most
notably Eu) can be detected in  lower $S/N$ data, and have led to
a number of stellar age measurements. Such ages can be derived from
a  ratio of a radioactive element to a stable
r-process nuclei (i.e. Th/r, U/r, U/Th; with r being stable elements
such as Eu, Os, and Ir), and comparing them with calculations of their
initial r-process productions (e.g. \citealt{schatz_chronometers}).
Since the giant CS~22892-052 is very C-rich, U could not be measured
but its Th/Eu ratio yielded an age of $14$\,Gyr
\citep{sneden03}. \citet{johnson_bolte} measured similar ages of five
 mildly r-enriched stars, and other studies have produced ages of
another few \citep{westin00, cowan_U_02,heresII,lai2008,hayek09}.

Compared to Th/Eu, the U/Th ratio is more robust to uncertainties in
the theoretically derived production ratio because Th and U haves
similar atomic masses (for which uncertainties would largely cancel
out; e.g., \citealt{wanajo2002, kratz2004}). Hence, stars displaying
Th \textit{and} U are the more desired r-process stars. For a similar
reason, stable elements of the 3$^{rd}$ r-process peak ($76 \le Z \le
78$) are best used in combination with the actinides (Th, $Z$\,=\,$90$
and U, $Z=92$).  The first U detection was made in the giant
CS~31082-001 \citep{Cayreletal:2001} giving an age of
$14$\,Gyr as derived from the U/Th ratio. Other chronometer
ratios in this star, such as Th/Eu, however, yielded
\textit{negative} ages. This is due to unusually high Th and U
abundances combined with an underabundance in Pb relative to the
majority of $r$-process enriched metal-poor stars. By now, there
are four known  r-process stars
\citep{Hilletal:2002,honda04,lai2008, hayek09} with such high Th/Eu
ratios ($\sim20\%$ of r-process stars). Since only the elements
heavier than the 3$^{\rm rd}$ $r$-process peak are (equally) affected
\citep{roederer09}, the U/Th ratio still gives a reasonable
age in CS31082-01. This behavior was termed an ``actinide-boost''
\citep{schatz_chronometers}, but it has become clear that these stars
likely have a different origin \citep{kratz07} than ``normal''
r-process stars. It also supports the conjecture that there are
multiple r-process sites \citep{Hilletal:2002}.

\begin{figure*}[!t]
\begin{center}
\includegraphics[width=16.8cm,clip=, bb=31 310 530 515]{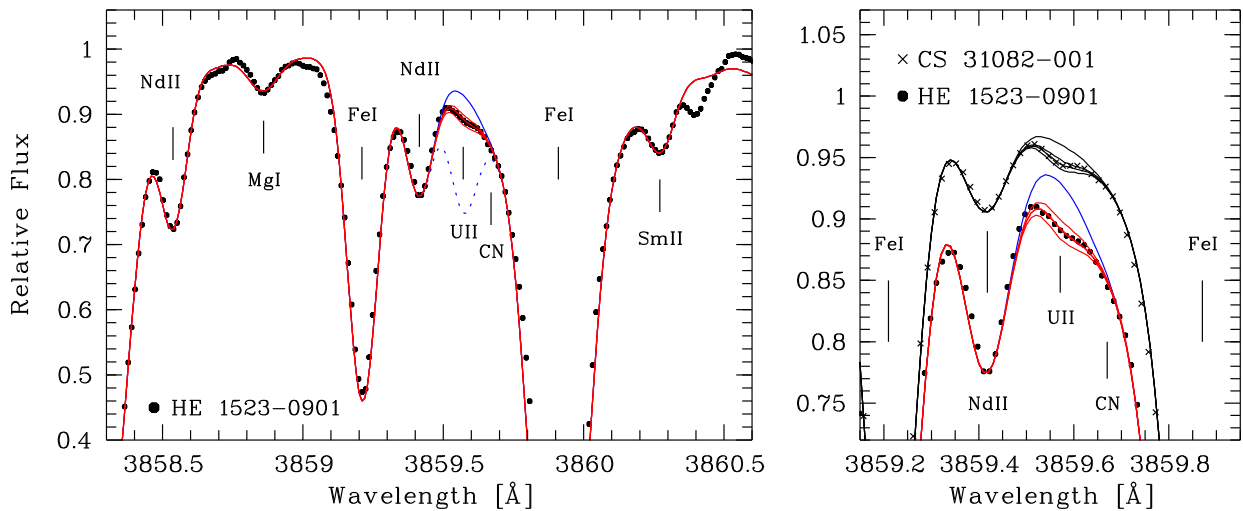}
  \caption{Spectral region around the \ion{U}{II}
  line in HE~1523$-$0901 (\textit{filled dots}) and CS~31082-001
  (\textit{crosses}; right panel only). Overplotted are synthetic
  spectra with different U abundances. The dotted line in the left
  panel corresponds to a scaled solar r-process U abundance present in
  the star if no U  decayed. Figure taken from \citet{he1523}.}\label{U_region} 
\end{center}
\end{figure*}

Only one star has so far been discovered for which measurements of Th
and U provide stellar age determinations from more than just one
chronometer ratio. The giant \linebreak HE~1523$-$0901 ($V = 11.1$) was found in
a sample of bright metal-poor stars \citep{frebel_bmps} from the
Hamburg/ESO Survey. It has the strongest enhancement in r-process
elements so far, $\mbox{[r/Fe]}=1.8$ \citep{he1523}, and among the
measured neutron-capture elements are Os, Ir, Th, and U.  In fact, the
U detection in this star is currently the most reliable one of only
\textit{three} stars with such measurements (in CS31082-01, and a
somewhat uncertain detection in BD +17$^{\circ}$ 3248;
\citealt{cowan_U_02}). Figure~\ref{U_region} shows the spectral region
around the U line from which the U abundance was deduced. The averaged
stellar age of $\sim13$\,Gyr \citep{he1523} is based on seven
chronometers Th/r, U/r and U/Th involving combinations of Eu, Os, Ir,
Th and U.

Unfortunately, realistic uncertainties for any such age measurements
range from $\sim2$ to $\sim5$\,Gyr (see \citealt{schatz_chronometers}
and \citealt{he1523} for a discussion).  Nevertheless, the stellar
ages of old stars naturally provide an important independent lower
limit to the age of the Galaxy, and hence, the Universe. The currently
accepted value for the age of the Universe is 13.7\,Gyr \citep{WMAP},
making these r-process stars some of the oldest known objects. This is
in line with the low metallicities of the stars that already indicates
a very early formation time.

In the absence of an age-metallicity relation for field halo stars,
the nucleo-chronometric ages thus demonstrate that metal-deficient
stars are indeed quite ancient, and well suited for studying the early
Universe.  By extension, this furthermore suggests that metal-poor
stars with similarly low Fe abundances but no excess in r-process
elements \linebreak should also be old. Moreover, the commonly made
assumption about the low mass (0.6 to 0.8\,$M_{\odot}$) of these
long-lived objects is justified as well.

\subsection{A Stellar Triumvirate: Th, U and Pb} 
Since different r-process models and the associated parameters usually
yield different r-process abundance distributions, particularly in the
heavy mass range, self-consistency constraints are very
valuable. Explaining the stellar abundance triumvirate of Th, U
\textit{and} Pb can provide such constraints. These three elements are
intimately coupled not only with each other but also to the conditions
(and potentially also the environment) of the r-process. Pb is the
$\beta$- plus $\alpha$-decay end-product of all decay chains in the
mass region between Pb and the onset of dominant spontaneous fission
above Th and U. It is also built up from the decay of Th and U
isotopes. Together with the Th and U measurements, a known Pb
abundance provides important constraints also on those poorly
understood decay channels. They are of critical importance for the
successful modeling of the r-process which, in turn, could provide
improved initial production ratios necessary for stellar age dating.

A new spectrum of HE~1523$-$0901 with $S/N\sim500$ was recently
obtained with VLT/UVES. Preliminary results indicate a detection of
this extremely difficult line measurement with an abundance of
$-0.4<\log\epsilon(\rm Pb)<-0.3$ (A. Frebel et al. 2010 in
preparation). To learn about the different contributions to the
production of lead, decay contributions of $^{238}{\rm U}$ into
$^{206}{\rm Pb}$, $^{232}{\rm U}$ into $^{208}{\rm Pb}$, and
$^{235}{\rm U}$ into $^{207}{\rm Pb}$ can be calculated (whereby the
last one is based on a theoretically derived ratio of $^{235}{\rm
U}/^{238}{\rm U}$). The total abundance of these three decays channels
amounts to \linebreak $\log\epsilon(\rm Pb)=-0.72$ which leaves ``room'' for the
direct and indirect decay channels that likely produce the main
portion of the Pb in the star.  Using r-process model calculations,
predictions were derived for the total Pb to be found in
HE~1523$-$0901. A site-independent model of the classical
``waiting-point'' approximation yielded $\epsilon (\rm{Pb})=-0.346$
\citep{frebel_ages} for a decay time of 13\,Gyr which is in agreement
with the preliminary abundances.  At the same time, this prediction is
less in agreement with the measured value of $\log\epsilon(\rm
Pb)=-0.55$ in CS31082-001 \citep{plez_lead}. This low Pb values is
difficult to understand theoretically (despite the actinide boost and
high Th and U), and is currently subject to much debate.

\subsection{The s-Process Signatures}\label{sec_s}

Neutron-capture elements are also produced in the interiors of low and
intermediate-mass asymptotic giant branch (AGB) stars through the
s-process and later dredged up to the surface. Unlike the
r-process, the s-process is not universal because two different sites
seem to host s-process nucleosynthesis.  The ``main'' component of the
s-process occurs in the helium shells of thermally pulsing lower mass
AGB stars and is believed to account for elements with $Z \ge 40$
(e.g., \citealt{arlandini1999, karakas02}). The so-called ``weak''
component occurs in the He- and C-burning cores of more massive stars,
and preferentially produces elements around $Z \sim 40$.

The s-process leads to a different characteristic abundance pattern of
neutron-capture elements than the \linebreak r-process. The signature is also
observed in some metal-poor stars, and their abundances follow that of
the scaled solar s-process. The s-process metal-poor stars received
their s-enriched material during a mass transfer event across a binary
system from a former more massive companion that went through its AGB
phase (e.g., \citealt{1997norriscarbon,2001aokisprocess, bisterzo09}).
During this period, large amounts of carbon are also transferred to
the companion producing a typical carbon-rich, s-rich metal-poor
stellar signature. Some metal-poor stars display signatures from both
r- and s-process enhancement in their abundance pattern (e.g.,
\citealt{cohen2003, barbuy2005, jonsell06, cash1}). Several different
scenarios have been invoked to explain the combination of the two
neutron-capture processes originating at two very different
astrophysical sites. However, no unambiguous explanation has yet been
found.

%%%%%%%%%%%%%%%%%%%%%%%%%%%%%%%%%%%%%%%%%%%%%%
\section{Tracing the Formation of the Galactic Halo with Metal-Poor Stars}

Simulations of the hierarchical assembly of galaxies within the cold
dark matter framework (\citealt{diemand07};\linebreak
\citealt{springel}) show that the Milky Way halo was successively
built up from small dark matter substructures, often referred to as
galactic building blocks, as long ago suggested by
\citet{sz78}. However, these simulations only include dark matter, and
it remains unclear to what extent small dark halos contain luminous
matter in the form of stars and gas. This question is particularly
important with respect to the so-called ``missing-satellite'' problem
which reflects the mismatch of the number of observed dwarf galaxies
surrounding the Milky with the predicted number of substructures for a
Milky Way-like halo. Studying the onset of star formation and
associated chemical evolution in dwarf galaxies thus provides some of
the currently missing information for our understanding of how the
observed properties of small satellites relate to the dark matter
substructures that build up larger galaxies.

\subsection{Chemical Evolution of Dwarf Galaxies and their Connection
to the Stellar Halo} 

The connection between the surviving dwarfs and those that dissolved
to form the halo is best addressed by examining in detail the stellar
chemical compositions of present-day dwarf galaxies. Establishing
detailed chemical histories of these different systems can provide
constraints on their dominant chemical enrichment events, as well as
the formation process of the Milky Way Specifically, detailed
knowledge of the most metal-poor (hence, oldest) stars in a given
system allow insight into the earliest phases of star formation before
the effects of internal chemical evolution were imprinted in stars
born later with higher metallicity (for further details on this topic
we refer the reader to the recent reviews by \citealt{tolstoy_araa}
and \citealt{koch_biermann}).

If the old, metal-poor halo was assembled from dwarf galaxies, the
metallicities of stars in dwarf galaxies must reach values
as low as (or lower) than what is currently found in the Galactic
halo. Moreover, the abundance ratio of those low-metallicity stars
must be roughly equal to those of equivalent stars in the halo.
Assuming that the currently observable dwarf galaxies are the
survivors of such an early accretion process, they provide an
opportunity to examine their stellar chemistry in search for such an
accretion process. Until recently, the common wisdom was, however,
that the ``classical'' dwarf galaxies in the Local Group (e.g.,
Carinae, Sextans, Sculptor, and Fornax), would not contain any stars
below $\mbox{[Fe/H]} = -3$ (e.g., \citealt{shetrone03}) despite the
fact that many halo stars exist with such low Fe values, and even some
with metallicities as low as $\mbox{[Fe/H]} \sim -5.0$.  Furthermore,
the higher-metallicity stars were found to have abundance ratios
different from comparable halo stars. Most strikingly, the
$\alpha$-element abundances are not enhanced, indicating different
enrichment mechanisms and timescales in these systems. This sparked a
debate about the viability of the \citet{sz78} paradigm, and in trying
to explain the origin of the metal-poor stellar halo.  However, it has
now been shown that this claim stems merely from biases in earlier
search techniques \citep{cohen09,kirby09, starkenburg10}. With
improved methods for identifying the lowest-metallicity objects (e.g.,
\citealt{kirby08}), extremely metal-poor stars with
$\mbox{[Fe/H]} < -3$ have already been identified in several dwarf
galaxies \citep{norris_boo, geha09, kirby08, ufs}.

\noindent
\textbf{Ultra-Faint Dwarf Galaxies}\\ \citet{kirby08} extended the
metallicity-luminosity relationship to the ultra-faint
($L<10^{5}\,L_{\odot}$) dwarfs and showed that the lowest luminosity
dwarfs have indeed the lowest average metallicities, with several
individual stars having $\mbox{[Fe/H]} < -3$.  The combined MDF of all
these systems goes down to $\mbox{[Fe/H]}\sim-3.3$, and the shape
appears similar to that of the Milky Way halo.  The ultra-faint dwarf
galaxies have large internal metallicity spreads, confirming 
earlier results at lower spectral resolution (e.g., \citealt{SG07},
\citealt{norris_boo}). They span more than $1$\,dex in $\mbox{[Fe/H]}$
in some of them. Such spreads indicate early star formation in
multiple proto-dwarf galaxies that later merged, extended star
formation histories, or incomplete mixing in the early ISM, or all of
the above.

The three brightest stars in each of Ursa Major\,II \linebreak
(UMa\,II) and Coma Berenices (ComBer) and two stars in Hercules are
the first stars in the ultra-faint dwarf galaxies to have been
observed with high-resolution spectroscopy. Two of them (in UMa\,II)
are also the first known extremely metal-poor stars which are not
members of the halo field population. Details on the observations and
analysis techniques are given in \citet{ufs} and \citet{koch_her}. For
the UMa\,II and ComBer stars, chemical abundances and upper limits of
up to 26 elements were determined for each star. The abundance results
demonstrate that the evolution of the elements in the ultra-faint
dwarfs is very similar to that of the Milky Way, and likely dominated
by massive stars. The $\alpha$-elements in these two ultra-faint dwarf
stars are overabundant, showing the halo-typical core-collapse SNe
signature. This is the first evidence that the abundance patterns of
light elements ($Z < 30$) in the ultra-faint dwarfs are remarkably
similar to the Milky Way halo. The agreement suggests that the
metal-poor end of the MW halo population could have been built up from
destroyed dwarf galaxies. Similar abundance results were also found by
other studies (\citealt{norris10,leo4, feltzing09}) but also
indications that the chemical evolution in these small systems may
have been very inhomogeneous.

The neutron-capture abundances are extremely low in the ultra-faint
dwarf stars. The observed Sr and Ba values are up to two orders of
magnitude below the abundances found in typical MW halo stars with
similar Fe content. However, a very large spread (up to 3\,dex) in
these elements is found among halo stars themselves. The large
depletions in the dwarf galaxy stars are thus not inconsistent with
the halo data since similarly low values are found in numerous 
objects. Interestingly though, the low neutron-capture abundances may
represent a typical signature of stars in dwarf galaxies. Similarly
low values are also found in Hercules \citep{koch_her} and Draco
\citep{fulbright_rich} despite their sometimes relatively high Fe
values of  $\mbox{[Fe/H]}\sim-2.0$ (in Hercules).

\vspace{0.2cm}
\noindent
\textbf{``Classical'' dSphs}\\ By applying the new search techniques
also to the more luminous dwarf galaxy Sculptor, the first extremely
metal-poor star in a classical dwarf was recently discovered (in a
sample of 380 stars; \citealt{kirby09}). The metallicity of
$\mbox{[Fe/H]} \sim -3.8$ was confirmed from a high-resolution
spectrum taken with Magellan/MIKE \citep{scl}. Only nine stars in the
halo have even lower Fe abundances than this object. This remarkable
finding suggests that more such low-metallicity stars could soon be
identified in the more luminous systems (see also
\citealt{starkenburg10}). The new star also shows that a metallicity
spread of $\sim3$\,dex is present in Sculptor. The chemical abundances
obtained from the high-resolution spectrum reveal a similar picture to
what has been found in the ultra-faint dwarf stars. The Sculptor star,
at $\mbox{[Fe/H]}\sim-3.8$, also has a remarkably similar chemical
make-up compared to that of the Milky Way halo at the lowest
metallicities. This is in contrast to what is found at higher
metallicities in these brighter dwarfs which have lower
$\alpha$-abundances than comparable halo stars (e.g.,
\citealt{shetrone03, geisler05}). There is increasing evidence, though
that a transition of halo-typical abundance ratios may take place
around a metallicity of $\mbox{[Fe/H]}= -3.0$ \citep{cohen09, aoki09}.

\subsection{The Origin of the Metal-Poor Tail of the Halo}
These new observational results are broadly consistent with the
predictions of currently favored cosmological models
(e.g. \citealt{robertson05}, \citealt{johnston08}). The majority of
the mass presumably in the inner part of the stellar halo (at
$\mbox{[Fe/H]}\sim-1.2$ to $-1.6$) was formed in much larger systems
such as the Magellanic Clouds. A scenario where the ultra-faint dwarf
galaxies contributed some individual metal-poor stars that are now
found primarily in the outer Galactic halo (although not exclusively)
is supported. However, these systems may not have been sufficiently
numerous to account for the entire metal-poor end of the Fe
metallicity distribution of the Milky Way halo. Since the classical
dSphs have more stellar mass and have been shown to also contain at
least some of the most metal-poor stars \citep{kirby09, scl}, they
could have been a major source of the lowest-metallicity halo stars.
Additional observations of more extremely metal-poor stars in the
various dwarf galaxies are highly desirable in the quest to determine
individual MDFs for each of these galaxies, and how those would
compare with each other and the Milky Way.

\section{Outlook -- What is Possible with Stellar Archaeology?}

Old metal-poor stars have long been used to learn about the conditions
of the early Universe.  This includes the origin and evolution of the
chemical elements, the relevant nucleosynthesis processes and sites
and the overall chemical and dynamical history of the Galaxy. By
extension, metal-poor stars provide constraints on the nature of the
first stars and their initial mass function, the chemical yields of
first/early SNe, as well as early star and galaxy formation processes
including the formation of the galactic halo.  Finding more of the
most metal-poor stars (e.g., stars with $\mbox{[Fe/H]}<-5.0$) would
enormously help to address all of these topics in more
detail. However, these stars are extremely rare \linebreak \citep{schoerck} and
difficult to find. The most promising way forward is to survey larger
volumes further out in the Galactic halo.

But how feasible is it to identify stars with even lower
metallicities?  \citet{poll} calculated the minimum observable Fe and
Mg abundances in the Galaxy by combining the critical metallicity of
$\mbox{[C/H]}_{min}=-3.5$ (the criterion for the formation of the
first low-mass stars by \citealt{brommnature}) with the maximum
carbon-to-iron ratio found in any metal-poor star. The resulting minimum Fe
value is $\mbox{[Fe/H]}_{min}=-7.3$. Analogously, the minimum Mg value \linebreak 
is $\mbox{[Mg/H]}_{min}=-5.5$. If $\mbox{[C/H]}_{min}$ was lower,
e.g., \linebreak $\mbox{[C/H]}_{min}=-4.5$, as suggested by recent dust cooling \linebreak 
computations, the minimum observable Fe and Mg abundances would
accordingly be lower. Spectrum synthesis calculations suggest these
low abundance levels are indeed \linebreak  measurable from each of the strongest Fe
and Mg lines in suitably cool metal-deficient giants.

Future surveys such as Skymapper and LAMOST will provide an abundance
of new metal-poor candidates as well as new faint dwarf galaxies. By
accessing such stars in the outer Galactic halo and dwarf galaxies we
will be able to gain a more complete census of the chemical and
dynamical history of our own Galaxy. Also, the lowest metallicity
stars are expected to be in the outer halo (e.g., \citealt{carollo}).
Their corresponding kinematic signatures may prevent them
from accreting too much enriched material from the ISM during their
lives so that their surface composition would not be altered (i.e.,
increased; \citealt{poll}). Hence, selecting for the most metal-poor
candidates will increasingly rely on our ability to combine chemical
abundance analyzes with kinematic information. Future missions such as
GAIA will provide accurate proper motions for \linebreak many object that have no
kinematic information available, including for most of the currently
known metal-poor giants.

However, many, if not most, of these future metal-poor candidates will
be too faint to be followed up with the high-resolution spectroscopy
necessary for detailed abundance analyzes. This is already an issue
for many current candidates leaving the outer halo a so far largely
unexplored territory: The limit for high-resolution work is
$V\sim19$\,mag, and one night's observing with 6-10\,m telescopes is
required for the minimum useful signal-to-noise ($S/N$) ratio of such
a spectrum. With the light-collecting power of the next generation of
optical telescopes, such as the Giant Magellan Telescope, the thirty
Meter Telescope or the European ELT, and if they are equipped with
high-resolution spectrographs, it would be possible to not only reach
out into the outer halo in search of the most metal-poor stars, but
also provide spectra with very high-$S/N$ ratio of somewhat brighter
stars. For example, r-process enhanced stars which provide crucial
empirical constraints on the nature of this nucleosynthesis process
require exquisite data quality e.g. for uranium and lead measurements
that are currently only possible for the very brightest stars.

It appears that the hunt for the metal-deficient stars in dwarf
galaxies may have just begun since these dwarfs host a large fraction
of low-metallicity stars, perhaps even much higher than what has so
far been inferred for the Galactic halo \citep{schoerck}. The detailed
abundance patterns of the stars in UMa\,II, ComBer, Leo\,IV and
Sculptor are strikingly similar to that of the Milky Way stellar halo,
thus renewing the support for dwarf galaxies as the building blocks of the
halo. Future discoveries of additional faint dwarf galaxies will
enable the identification of many more metal-poor stars in new,
low-luminosity systems.  But also the brighter dSphs have to be
revisited for their metal-poor content \citep{kirby09}.  More stars at
the lowest metallicities are clearly desired to better quantify the
emerging chemical signatures and to solidify our understanding of the
early Galaxy assembly process. Together with advances in the
theoretical understanding of early star and galaxy formation and SNe yields, a
more complete picture of the evolution of the Milky Way Galaxy and
other systems may soon be obtained. Only in this way can the
hierarchical merging paradigm for the formation of the Milky Way be
put on firm observational ground.

\acknowledgements I am very grateful to the Astronomische\linebreak
Gesellschaft and its selection committee for awarding me the 2009
Ludwig-Biermann Award. Lars Hernquist and Ian Roederer have given
useful comments to the manuscript, and John Norris provided the
spectrum of G66-30 shown in Figure~1. I warmly thank my many wonderful
collaborators who have always inspired me, and make working in this
field a great pleasure: Wako Aoki, Martin Asplund, Tim Beers, Volker
Bromm, Norbert Christlieb, Karl-Ludwig Kratz, Evan Kirby, John Norris,
Ian Roederer, Josh Simon, Chris Sneden and many others. This work has
been supported by a Clay Postdoctoral Fellowship administered by the
Smithsonian Astrophysical Observatory.

\section*{Appendix}
References for the collection of the literature data.
%refs for master_heavy_ncap.txt
\citet{aoki_lead2000},  
\citet{2001aokisprocess},  
\citet{aoki_lead2002}, 
\citet{aoki2002_lp625}, 
\citet{aoki_cempno}, 
\citet{aoki_mg},     
\citet{aoki05},
\citet{aoki2006_osir}, 
\citet{aoki_cemp_2007}, 
\citet{aoki2007_cos82}, 
\citet{aoki_studiesIV}, 
\citet{aoki2008}, 
\citet{aoki09}, 
\citet{Aokihe1327}, 
\citet{arnone}, 
\citet{heresII}, 
\citet{barbuy2005}, 
\citet{bonifacio09}, 
\citet{2000burris}, 
\citet{carretta02}, 
\citet{heresI}, 
\citet{cohen2003}, 
\citet{cohen04}, 
\citet{cohen2006}, 
\citet{cohen_highSi}, 
\citet{cohen08}, 
\citet{cohen09}, 
\citet{collet06}, 
\citet{cowan_U_02}, 
\citet{francois07}, 
\citet{frebel_he1300}, 
\citet{he1327_uves}, 
\citet{ufs}, 
\citet{scl}, 
\citet{fulbright}, 
\citet{fulbright_rich}, 
\citet{feltzing09}, 
\citet{geisler05}, 
\citet{hayek09}, 
\citet{Hilletal:2002}, 
\citet{plez_lead}, 
\citet{honda04}, 
\citet{honda06}, 
\citet{honda07}, 
\citet{ito2009}, 
\citet{ivans_alphapoor}, 
\citet{ivans05}, 
\citet{ivans06}, 
\citet{johnson2002_cs22183_015}, % xx check how man JOH02 are there..
\citet{johnson2002_rproc}, \linebreak 
\citet{johnson2002_23stars},
\citet{johnson_bolte}, 
\citet{johnson2004},
\citet{aoki2006_osir}, 
\citet{jonsell05},  
\citet{jonsell06}, 
\citet{koch_her}, 
\citet{lai2007}, 
\citet{lai2008}, 
\citet{lai2009}, 
\citet{lucatello03}, 
\citet{masseron2006}, 
\citet{McWilliametal,mcwilliam98}, 
\citet{norris2000}, 
\citet{Norrisetal:2001}, 
\citet{Norrisetal:2002},  
\citet{he0557}, 
\citet{norris_boo}, 
\citet{norris_cempno}, 	 
\citet{1997norriscarbon}, 
\citet{norris_emp3},    
\citet{preston_sneden01}, 
\citet{preston_bmp}, 
\citet{preston_rhb}, 
\citet{cash1}, 
\citet{Ryanetal:1991}, 
\citet{ryan96}, 
\citet{shetrone01}, 
\citet{shetrone03}, 
\citet{leo4}, 
\citet{sivarani04}, 
\citet{sivarani06}, 
\citet{sneden03}, 
\citet{spite2005}, 
\citet{spite2000}, 
\citet{westin00}, 
\citet{cayrel2004}, 
\citet{zacs}

\end{document}